\documentclass[a4paper,12pt]{article}
\usepackage{graphicx}
\usepackage{epsfig}
\usepackage{jcappub}

\begin{document}

\title{Growth Diagnostics for Dark Energy models and EUCLID forecast}
\author{Sampurnanand $^{1,2}$, Anjan A Sen $^1$}
\affiliation{$^1$Center For Theoretical Physics, Jamia Millia Islamia, New Delhi-110025, India}
\affiliation{$^2$Department of Physics and Astrophysics, University of Delhi, Delhi-110007, India}

\emailAdd{sampurna@physics.du.ac.in,  aasen@jmi.ac.in}

\abstract{In this work we introduce a new set of parameters $(r_{g}, s_{g})$ involving the linear growth of matter perturbation that can distinguish and constrain different dark energy models very efficiently. Interestingly, for $\Lambda$CDM model these parameters take exact value $(1,1)$ at all red shifts whereas for models different from $\Lambda$CDM, they follow different trajectories in the $(r_{g}, s_{g})$ phase plane. By considering the parametrization for the dark energy equation of state ($w$) and for the linear growth rate ($f_{g}$), we show that different dark energy behaviours with similar evolution of the linear density contrast, can produce distinguishable trajectories in the $(r_{g}, s_{g})$ phase plane. Moreover, one can  put stringent constraint on these phase plane using future measurements like EUCLID ruling out some of the dark energy behaviours.}
\date{today}

\maketitle

\section{Introduction}
\noindent
One of the biggest discoveries in modern cosmology is the fact that our Universe is currently undergoing an accelerated expansion phase and the change over from decelerated phase has occurred in recent past. Various cosmological observations like luminosity distance measurements from  Type-Ia Supernovae \cite{Riess98}-\cite{panstarrs1}, measurement of angular diameter distance using standard ruler like acoustic oscillations in Cosmic Microwave Background Radiation (CMBR) \cite{cmb} as well as Baryon Acoustic Oscillations (BAO)\cite{bao} in matter power spectra and measurements of gravitational clustering \cite{Eisenstein} suggest that two third of the total energy density of our Universe is contributed by an exotic component with negative pressure (known as { \it dark energy}) which is responsible for the late time acceleration of the Universe.  Till date, a variety of dark energy models have been proposed involving cosmological constant, canonical and  non-canonical scalar fields, Galileon fields, DBI-Galileon fields, phantom fields, scalar fields non-minimally coupled to gravity, chaplygin and generalized chaplygin gas, fluids involving defects and many more, to explain the late time acceleration of the Universe (see \cite{derev} and references therein). Models involving infra-red modification of the gravity sector have also been proposed to explain this (see \cite{fr} for an excellent review on $f(R)$ gravity models). The challenge now is to distinguish and constrain these models using present and future cosmological observations. If we rely only on the measurements involving background expansion, it is very difficult to remove the degeneracies among  different models. In this regard, two very important diagnostics have been proposed by Sahni et. al.\cite{Sahni03}. These are called {\it statefinders} $(r, s)$. These two parameters are so constructed that they take fixed values $(1,0)$ at all red shifts for $\Lambda$CDM model. If one studies the phase space in the $(r, s)$ plane, $\Lambda$CDM represents a fixed point and all other dark energy models show widely different trajectories. Hence constraining the $(r, s)$ phase plane can remove degeneracies between different dark energy models including the concordance $\Lambda$CDM model. It was also shown by the same authors \cite{Alam}  that future Type-Ia supernova observation like SNAP and JDEM can constraint this phase space so severely that many of the possible dark energy models can be ruled out.

The other important probe for the late time cosmic acceleration is the growth of the matter density contrast. Cosmic acceleration slows down the growth of the density fluctuations and hence its effect on the matter density contrast $\delta_{m}(z) = \frac{\delta\rho_{m}}{\rho_{m}}$ can be a very important tool to constrain different dark energy models. It can provide crucial insight into the dark energy properties to remove the degeneracies involved. We construct a set of new parameters to demonstrate the importance of the growth of matter density contrast to distinguish various class of dark energy models.

This paper is organized as follows: in Section 2, we introduce the two newly constructed parameters  and demonstrate their effectiveness. In Section 3, we forecast the constraints on these parameters with future measurement like EUCLID. Finally, we conclude in section 4.

\section{The Growth Diagnostics} 

\noindent
\begin{figure}
\begin{center}
\begin{tabular}{|c|c|}
\hline
{\includegraphics[width=3.2in,height=2.6in,angle=0]{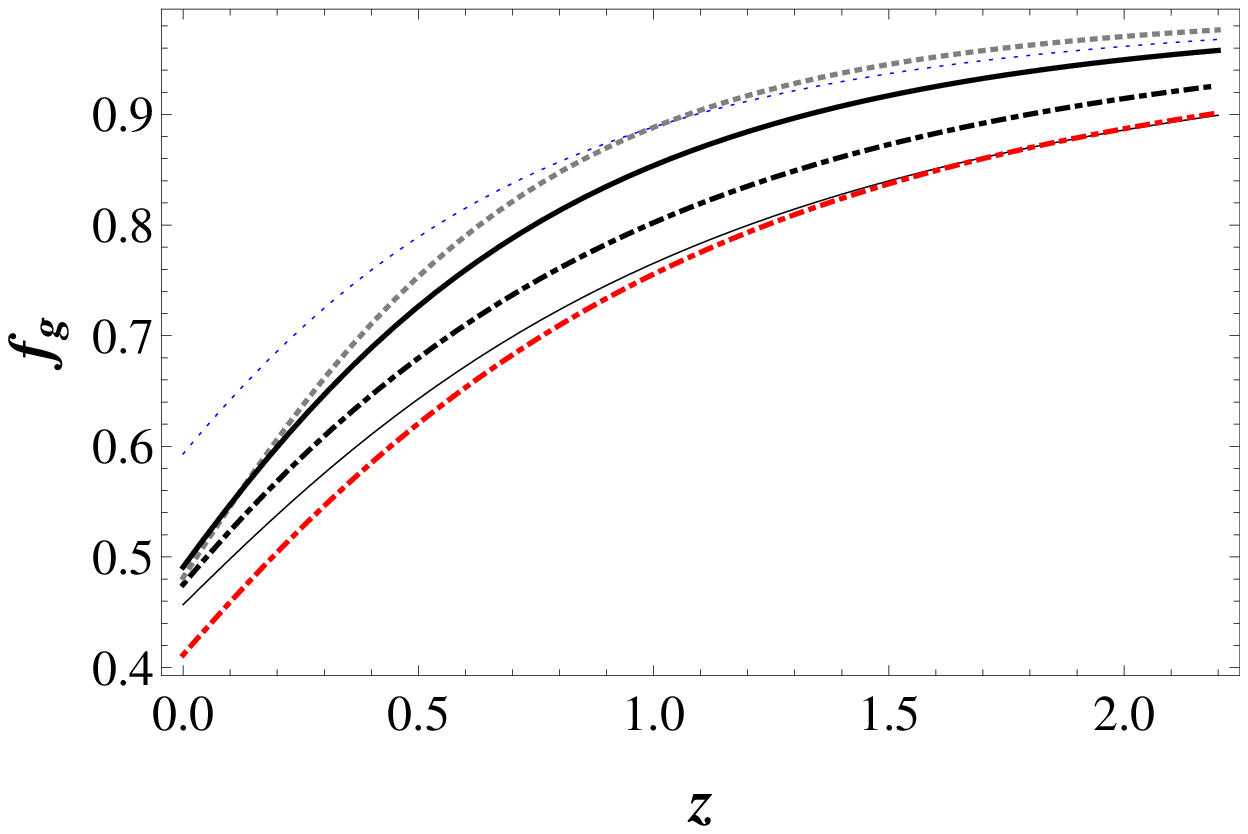}}&
{\includegraphics[width=2.8in,height=2.6in,angle=0]{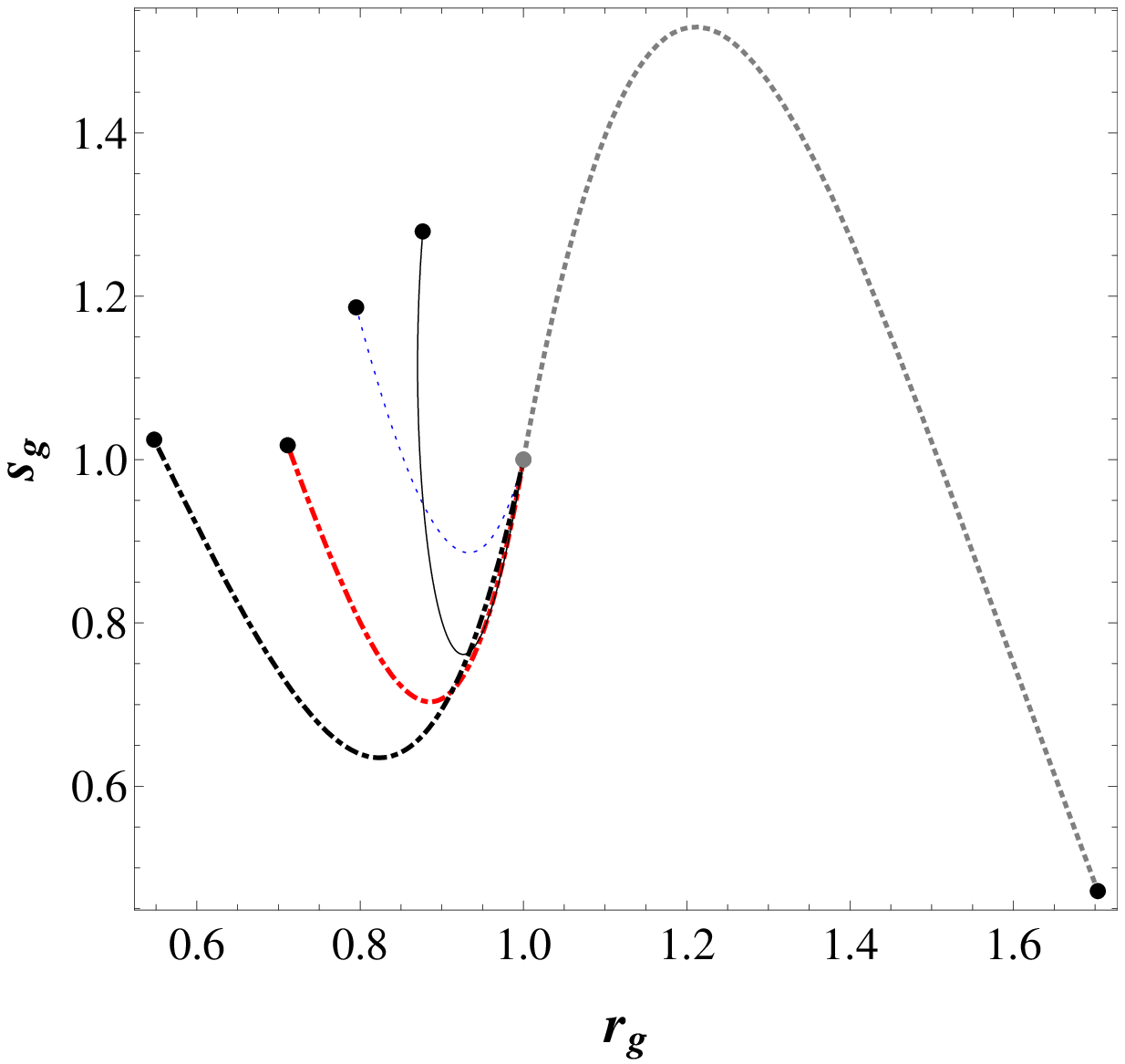}}\\
\hline
 \end{tabular}
\caption{Behaviour of growth factor $f_{g}$ as a function of red shift $z$(left). Evolution of various models in the $r_{g} - s_{g}$ phase space (Right). In the left panel, {\it Black thick solid} line denotes the $\Lambda$CDM model with $w_{0}= -1, w_{a}= 0, \gamma_{0}=0.545, \gamma_{a}=0$. The same model is represented by gray dot in the right panel. In both the figures, {\it Red thick dot dashed line} denotes the DGP model for which $w_{0} = -0.8, w_{a} = 0.0, \gamma_{0}=0.68, \gamma_{a}=0.0$, {\it Blue dotted line} represents f(R) model for which $w_{0} = -0.99, w_{a} = 0, \gamma_{0}=0.4, \gamma_{a} = 0$ , {\it  dot dashed line} denotes $w_{0} = -0.9, w_{a} = 0.2, \gamma_{0}=0.6, \gamma_{a} = 0.2$, {\it Black thick dot dashed line} denotes gcg non-phantom model for which $w_{0} = -0.75, w_{a} =-0.12, \gamma_{0}=0.57, \gamma_{a} = -0.1$ and {\it Gray dotted line} denotes gcg phantom for which $w_{0} = -1.22, w_{a} = 0.01, \gamma_{0}=0.56, \gamma_{a} = 0.1.$} \label{rgsg}
\end{center}
\end{figure}
We introduce two new dimensionless parameters $r_{g}$ ans $s_{g}$ which are constructed using matter overdensity and its derivatives. 
We call them ”Growth Diagnostics”. These parameters are defined as:
\begin{eqnarray}
r_{g}& = & \frac{3}{2}\Omega_{m} - \frac{\ddot{F}}{2H\dot{F}} , \label{param} \\
s_{g}& = &\frac{3}{4}(3\Omega_{m}-2)^{2} - \frac{\dddot{F}}{2H^{2}\dot{F}} .  \nonumber 
\end{eqnarray}
where
$F(t)$ is a dimensionless quantity defined as:
\begin{equation}
F(t)= \frac{\delta_{m}}{h}  \label{def1} ,
\end{equation}
and $\delta_{m}$ is the linear matter density contrast satisfying the equation:

\begin{equation}
\ddot{\delta_{m}} + 2H\dot{\delta_{m}} = 4\pi G \rho_{m}\delta_{m}. \label{denevoleqn}
\end{equation}
 Here $H = \frac{\dot{a}}{a}$ is the Hubble parameter, $a$ is the scale factor, and  $\rho_{m}$ is the background matter density. The dot denotes the derivative with respect to cosmic time $t$. $h= \frac{H}{H_{0}}$ is the normalized Hubble's constant. $\Omega_{m}$ is the matter density parameter $(\rho_{m}/\rho_{c})$. Growth rate is defined as $f_{g}= \frac{d log \delta_{m}}{d log a}$. Since the parameters defined in Eq.(\ref{param}) involves $\delta_{m}$ and its time derivative, so it can be rewritten in terms of $f_{g}$ as well. The growth of the large scale structures are measured through the matter density contrast at linear regime and sub horizon scales. On these scales, it is safe to assume the dark energy to be homogeneous fluid. Hence in Eq.(\ref{denevoleqn}), we ignore the dark energy perturbation. \\
\noindent
These new parameters are defined such that both of them are unity for $\Lambda$CDM at all red shifts. Therefore, in $(r_{g}, s_{g})$ phase plane, $\Lambda$CDM represents a fixed point (1,1). For all other dark energy models we get deviation from the fixed point. We shall now demonstrate the potential of the parameter set to remove the degeneracies among various dark energy models having similar growth rate today. \\
\noindent
In this work, we consider following parametrizations:
\begin{itemize}
\item  CPL parametrization \cite{Chevallier, Linder} for dark energy equation of state:
\begin{equation}
w  =  w_{0} + w_{a}(1-a),
\end{equation}
where $w_{0}$ and $w_{a}$  are constants. In this parametrization, $w_{a} = 0$ corresponds to the constant equation of state. Therefore, $w_{0}= -1 $ and $w_{a}= 0$ will represent $\Lambda$CDM model.
\item $\gamma$- parametrization for the growth rate \cite{Gannouji,wu,Fu}: 
\begin{equation}
f_{g}(a) = \Omega_{m}(a)^{\gamma(a)} ,\label{growth_rate}
\end{equation}
where,
\begin{equation}
\gamma  =  \gamma_{0} + \gamma_{a}(1-a). \label{gr_index}
\end{equation}
$\gamma_{0}$, $\gamma_{a}$  are constants and $\Lambda$CDM model is reproduced for $\gamma_{0} = 0.545$ and $\gamma_{a} = 0$ in this parametrization.\\
\noindent
We assume three different kind models for explaining late time acceleration: 

\noindent
 \item {\bf DGP model}: For the flat space considered by Marteens and Majerotto\cite{Marteens}, the DGP model is given by $w_{0}=-0.8, w_{a}=0$ and $\gamma_{0} = 0.68, \gamma_{a}=0$ \cite{linder_cahn}.
\item {\bf f(R) model}: The Hu and Sawiki\cite{HU_sawicki} model of f(R) is given by $w_{0}= -0.99, w_{a}=0$ and $\gamma_{0} = 0.4, \gamma_{a}=0$.
\item {\bf Generalized Chaplygin Gas (GCG)}: GCG \cite{sen} is characterized by the equation of state
\begin{equation}
w_{gcg} = -\frac{A_{s}}{A_{s}+(1-A_{s})(1+z)^{3(1+\beta)}}
\end{equation}
where $A_{s} $ and $\beta$ are constants. $A_{s} > 0$ and $z=0, A_{s}=1$ represents $\Lambda$CDM.\\
\begin{enumerate}
\item $0< A_{s} < 1$ and $\beta > -1$ corresponds to the tracking type scalar field models.
\item $0< A_{s} < 1$ and $\beta < -1$ corresponds to the thawing type scalar field models.
\item $ A_{s} > 1$ and $\beta > -1$ corresponds to the phantom models.
\end{enumerate}
\end{itemize}
\noindent 
Left panel of Fig.(\ref{rgsg}) shows the variation of $f_{g}$ with red shift for different models characterized by different choices of $(w_{0}, w_{a}, \gamma_{0}, \gamma_{a)}$. Evolution of the same models in the $r_{g}-s_{g}$ phase plane are shown in the right panel of the same figure. Black dots represents the present values of $(r_{g},s_{g})$. The $\Lambda$CDM model is represented by fixed point. It is demonstrated that different dark energy models which produces nearly similar behaviour for the growth rate $f_{g}(a)$, have different trajectories in $r_{g}-s_{g}$ phase plane. It is also shown that the DGP and modified gravity model are more distinguished in the $r_{g}-s_{g}$ phase plane. The values of  $r_{g}$ and $s_{g}$ at present are also widely spread for different dark energy models. This shows that these two growth diagnostics $(r_{g}, s_{g})$ can be very useful in distinguishing different dark energy models.  

\section{Constraints with EUCLID forecast}

After demonstrating the interesting features of the growth diagnostics in the previous section, our goal is to  put constraints on these parameters with future experiments like EUCLID. In order to do that, we use the reference {\it pseudo $\Lambda$CDM} model\cite{Amendola} characterized by $w_{0} = -0.95$ and $w_{a}=0$. Growth rate in this case is obtained from Eq.(\ref{growth_rate}) by setting  $\gamma_{0} = 0.545$ and $\gamma_{a}=0$ in Eq.(\ref{gr_index}). The evolution of $\Omega_{m}$ is given as:
\begin{equation}
\Omega_{m}(a) = \frac{\Omega_{m0}a^{-3}}{h^{2}}, \nonumber
\end{equation}
where,
\begin{equation}
h^{2} = \Omega_{m0}a^{-3} + (1-\Omega_{m0})a^{-3(1+w_{0}+w_{a})}e^{-3w_{a}(1-a)}.
\end{equation}
We assume flat universe with $\Omega_{k} = 0$ and $\Omega_{m0} = 0.27$. The fiducial value of growth rate at different red shifts along with errors are given in Table(\ref{fdata}), taken from  \cite{Amendola} .\\
Our analysis is done as follows:
\begin{itemize}
\item  $w_{0}, w_{a}$ fixed at values for reference model :\\
With the dataset given in Table(\ref{fdata}), we first obtain the Fisher Matrix for $\gamma_{0}, \gamma_{a}$ keeping  $w_{0}, w_{a}$ fixed at values for reference model. Since $r_{g}$ and $s_{g}$ are known functions of $\gamma_{0}, \gamma_{a}$, we can calculate the Fisher matrix for $r_{g0}, s_{g0}$ (subscript '0' means $z=0$), given the Fisher Matrix for $\gamma_{0}, \gamma_{a}$ using the relation\cite{Dan}:
\begin{equation}
[{\cal F'}] = [{\cal M}^{T}][{\cal F}][{\cal M}]  \label{FisherMatrixRel}
\end{equation}

where, 
\begin{equation}
{\cal M}  =  \left(
           \begin{array}{cc}
            \frac{\partial \gamma_{0}}{\partial r_{g0}} &  \frac{\partial \gamma_{0}}{\partial s_{g0}}\\
              \frac{\partial \gamma_{a}}{\partial r_{g0}} & \frac{\partial \gamma_{a}}{\partial s_{g0}}\\
           \end{array}
           \right)_{Best Fit}
\end{equation}

\item  $\gamma_{0}, \gamma_{a}$ fixed at values for reference model :\\
\noindent In this case we keep $ \gamma_{0}, \gamma_{a} $ fixed at values for reference model and obtain the Fisher matrix for $w_{0}, w_{a}$ using data given in Table(\ref{fdata}). Knowing the Fisher Matrix for $w_{0}, w_{a}$,  Fisher matrix for $r_{g0}-s_{g0}$,  can be computed using the relation:

\begin{equation}
[{\cal F'}] = [{\cal M}^{T}][{\cal F}][{\cal M}]  \label{FisherMatrixRel}
\end{equation}

where, 
\begin{equation}
{\cal M}  =  \left(
           \begin{array}{cc}
            \frac{\partial w_{0}}{\partial r_{g0}} &  \frac{\partial w_{0}}{\partial s_{g0}}\\
              \frac{\partial w_{a}}{\partial r_{g0}} & \frac{\partial w_{a}}{\partial s_{g0}}\\
           \end{array}
           \right)_{Best Fit}
\end{equation}
\end{itemize}

\begin{figure}
\begin{center}
\begin{tabular}{|c|c|}
\hline
{\includegraphics[width=2.9in,height=3in,angle=0]{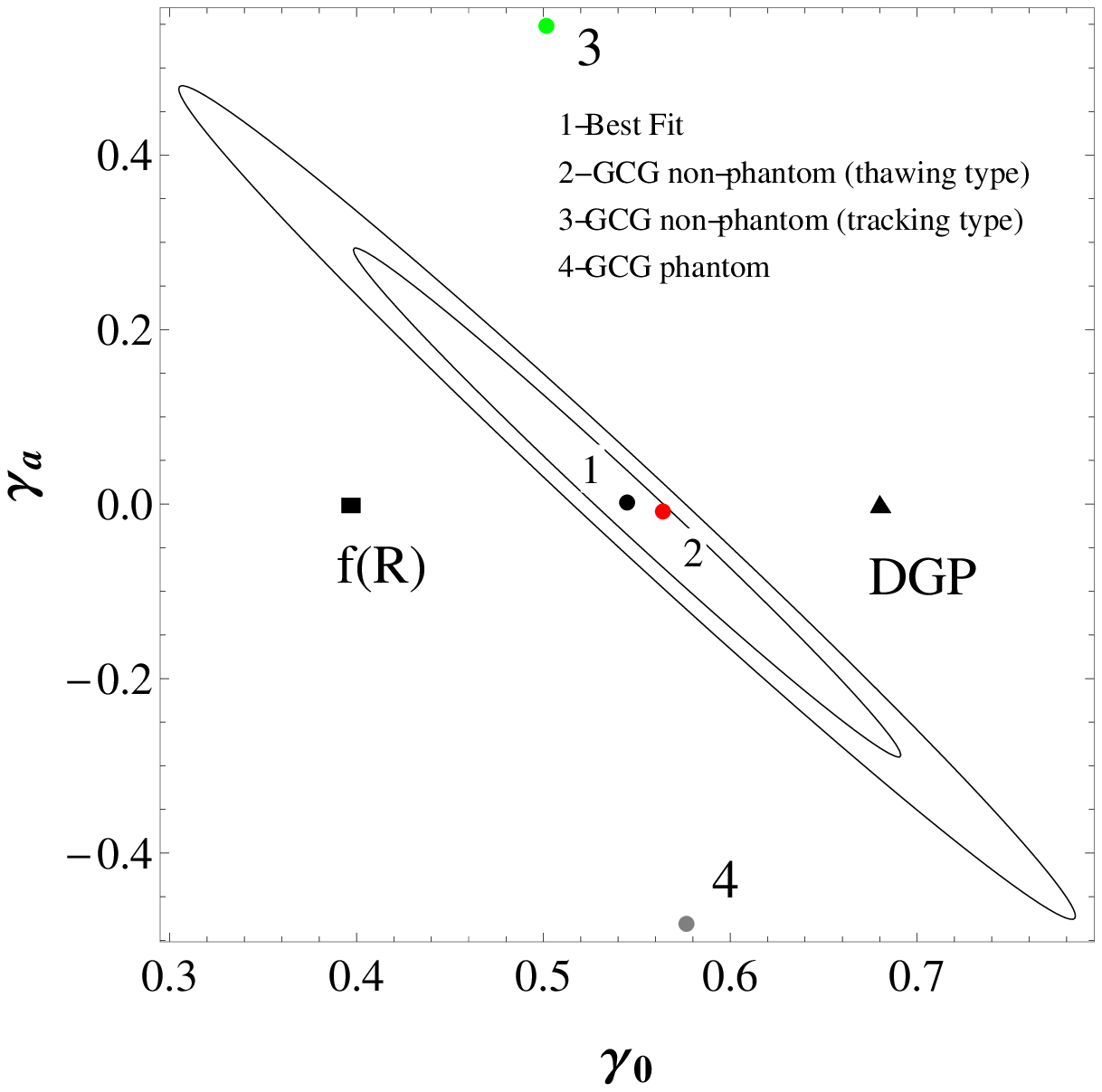}}&
{\includegraphics[width=2.8in,height=3in,angle=0]{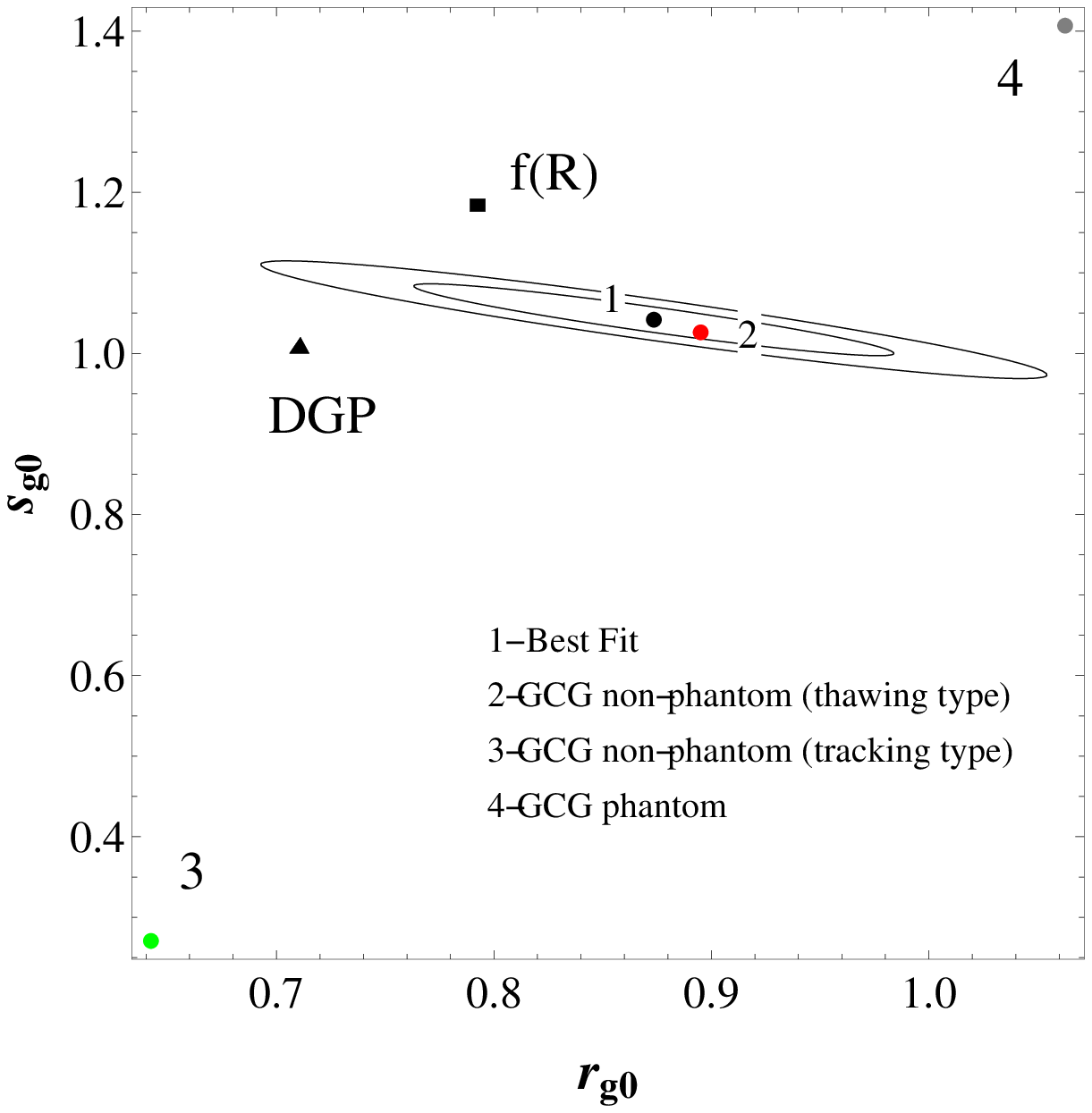}}\\
\hline
 \end{tabular}
\caption{ $1-\sigma$ and $2-\sigma$ contours in the $\gamma_{0}-\gamma_{a}$ plane. (left).  $1-\sigma$ and $2-\sigma$ contours in the $r_{g0}-s_{g0}$ plane. (right)} \label{g_index_contour}
\end{center}
\end{figure}

\begin{figure}
\begin{center}
\begin{tabular}{|c|c|}
\hline
{\includegraphics[width=2.9in,height=3in,angle=0]{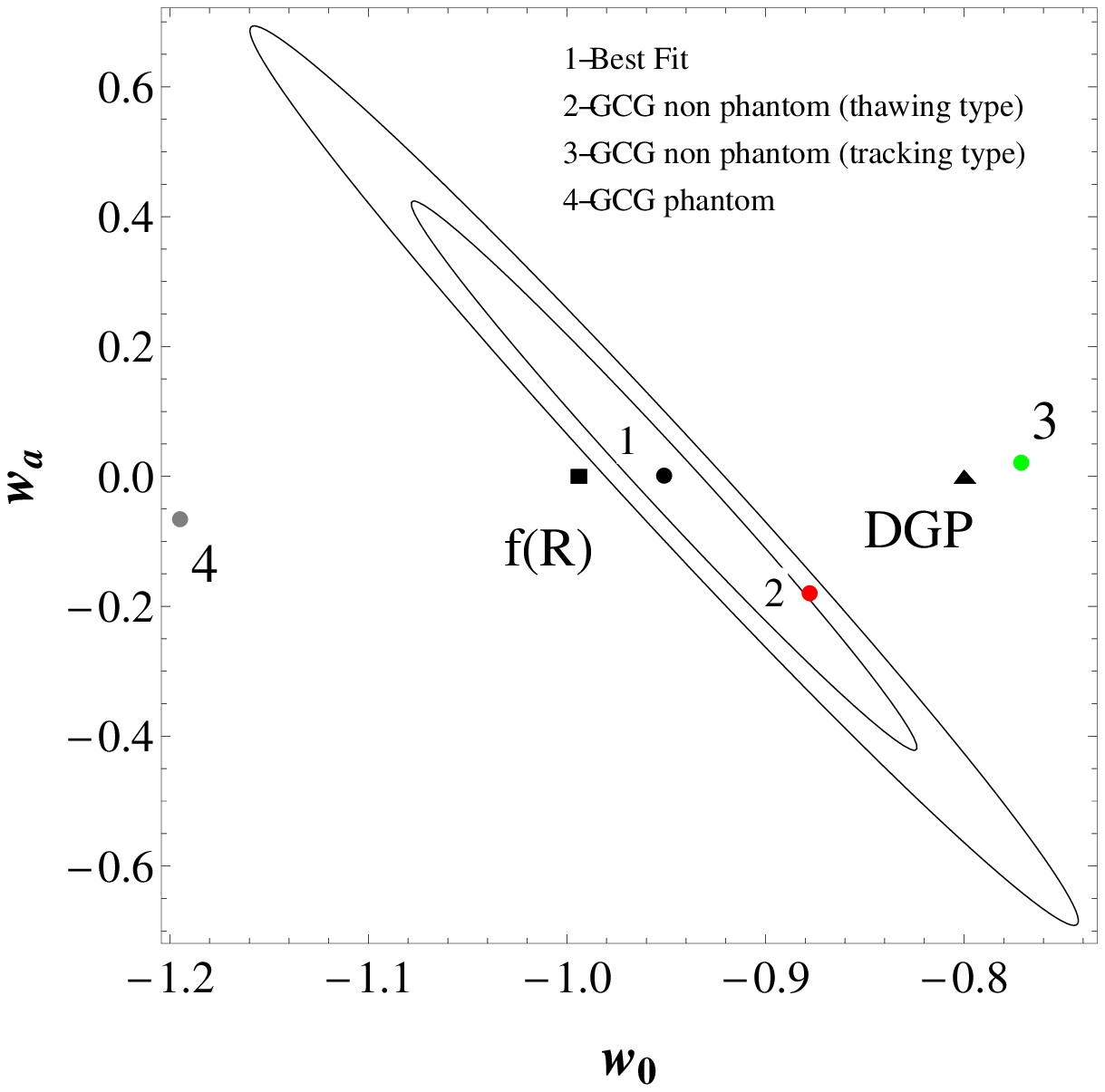}}&
{\includegraphics[width=2.8in,height=3in,angle=0]{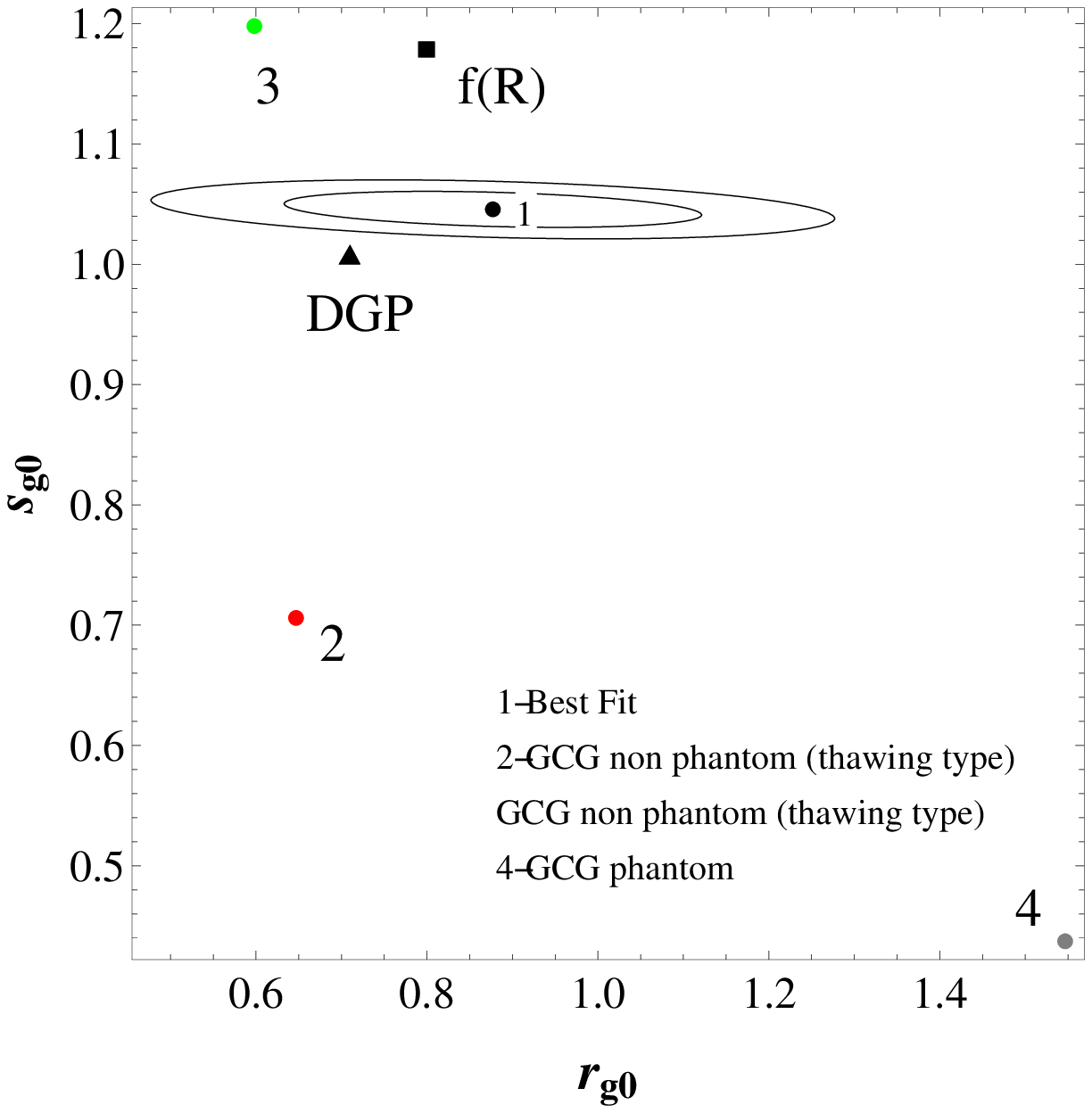}}\\
\hline
 \end{tabular}
\caption{ $1-\sigma$ and $2-\sigma$ contours in the $w_{0}-w_{a}$ plane. (left) and corresponding $1-\sigma$ and $2-\sigma$ contours in the $r_{g0}-s_{g0}$ plane. (right)} \label{eos_contour}
\end{center}
\end{figure}

\begin{table}[h!]
\begin{center}
\vspace{6pt}
\begin{tabular}{lllll }
\hline
\\
$z$ & & $f_{g}$ & & $\sigma$ \\
\\
\hline \\
0.7 &\hspace{3cm} &0.76 &\hspace{3cm} &0.011  \\
0.8 &\hspace{3cm} &0.80 &\hspace{3cm} &0.010  \\
0.9 &\hspace{3cm} &0.82 &\hspace{3cm} &0.009  \\
1.0 &\hspace{3cm} &0.84 &\hspace{3cm} &0.009 \\
1.1 &\hspace{3cm} &0.86 &\hspace{3cm} &0.009   \\
1.2 &\hspace{3cm} &0.87 &\hspace{3cm} &0.009   \\
1.3 &\hspace{3cm} &0.88 &\hspace{3cm} &0.010  \\
1.4 &\hspace{3cm} &0.89 &\hspace{3cm} &0.010   \\
1.5 &\hspace{3cm} &0.91 &\hspace{3cm} &0.011   \\
1.6 &\hspace{3cm} &0.91 &\hspace{3cm} &0.012   \\
1.7 &\hspace{3cm} &0.92 &\hspace{3cm} &0.014   \\
1.8 &\hspace{3cm} &0.93 &\hspace{3cm} &0.014 \\
1.9 &\hspace{3cm} &0.93 &\hspace{3cm} &0.017   \\
2.0 & \hspace{3cm}&0.94 &\hspace{3cm} &0.023   \\
\\
\hline
\end{tabular}
\caption{Fiducial values for linear growth rate $f_{g}$ along with errors, taken from \cite{Amendola} \label{fdata}}
\end{center}
\end{table}
\noindent
Knowing the Fisher matrices (and hence covariance matrices), we can draw the confidence contours. In Fig.(\ref{g_index_contour}) we have drawn the $1\sigma$ and $2\sigma$ confidence contours in the $\gamma_{0}-\gamma_{a}$ parameter space (left panel) and the corresponding contours in the $r_{g0}-s_{g0}$ phase space (right panel). \\
The $1\sigma$ and $2\sigma$ confidence contours in the $w_{0}-w_{a}$ parameter space  and the corresponding contours in the $r_{g0}-s_{g0}$ phase space are shown in Fig.(\ref{eos_contour}).\\
\noindent

The second case, shown in the  Fig.(\ref{eos_contour}) is particularly interesting.  It shows that GCG non-phantom thawing model is allowed in the $w_{0}-w_{a}$ plane, whereas in the $rg_{0}-sg_{0}$ plane, it is ruled out with very high confidence limit. Similarly the $f(R)$ model which was at the boundary of the $2\sigma$ confidence limit in the $w_{0}-w_{a}$ plane is now disallowd with more cofidence limit. This shows that models which are apparently consistent with the observational data in the $w_{0}-w_{a}$ plane, can be ruled out using the $rg_{0}-sg_{0}$ plane. Hence for growth measurments, $rg_{0}-sg_{0}$ plane can be more interesting to study.

We now calculate the Figure of Merit (FoM) to quantify how constraining $r_{g0}-s_{g0}$ plane is compared to $\gamma_{0}-\gamma_{a}$ and $w_{0}-w_{a}$ plane. Table(\ref{g_index_fom}) compares the figure of merits for both the cases. We observe that the FoM for $r_{g0}-s_{g0}$ is almost three time the FoM for $\gamma_{0}, \gamma_{a}$ and $w_{0}-w_{a}$. Consequently,  $r_{g0}-s_{g0}$ parameter space is more constraining than the  $\gamma_{0}, \gamma_{a}$ or $w_{0}-w_{a}$ plane. As a result, $r_{g0}-s_{g0}$ can be used to constrain the dark energy models more efficiently.\\

\begin{table}[h!]
\begin{center}
\begin{tabular}{llllll}
\hline 
\\
&  & FOM for $\gamma_{0}-\gamma_{a}$ & & FOM for $ r_{g0}-s_{g0}$ & 
\\\\
\hline
\\
 keeping                      & \hspace{1cm} &          &  \hspace{1cm}  & \\
 $w_{0}, w_{a}$     & \hspace{1cm} & 426.547  & \hspace{1cm}   & 1316.08 \\
fixed                         & \hspace{1cm} &          &  \hspace{1cm}  &  \\\\
\hline 
\\
&  & FOM for $w_{0}-w_{a}$ & & FOM for $ r_{g0}-s_{g0}$ & \\
\\
\hline
\\
 keeping                      & \hspace{1cm} &          &  \hspace{1cm}  & \\
 $\gamma_{0}, \gamma_{a}$     & \hspace{1cm} & 297.107  & \hspace{1cm}   & 662.143 \\
fixed                         & \hspace{1cm} &          &  \hspace{1cm}  &  \\\\
\hline 
\end{tabular}
\caption{Figure of merit for the contours in Fig(\ref{g_index_contour}) and Fig(\ref{eos_contour}) }
\label{g_index_fom}
\end{center}
\end{table}
\noindent
However, it may happen that the distances between various dark energy models and the best fit point (which corresponds to the reference pseudo-$\Lambda$CDM model) have also reduced in the $r_{g0}-s_{g0}$ phase plane. In that case, the reduction of the size of the contours in the $r_{go}-s_{g0}$ plane may not be constraining enough. To check this, we have calculated these distances for both the cases considered above and is mentioned in Table(\ref{dist1}) and Table(\ref{dist2}).\\ 
\begin{table}[]
\begin{center}
\vspace{6pt}
\begin{tabular}{l|ccccc}
\hline\\
&  f(R) & DGP  & GCG-thawing  & GCG tracking  & GCG phantom  \\ \\
\hline \\
$\gamma_{0}-\gamma_{a}$ plane& 0.145 &0.135  &0.022 &0.548 &0.484  \\ \\
\hline \\
$r_{g0}-s_{g0}$ plane&0.165 &0.164  &0.027 &0.805 &0.411 \\
\\
\hline
\end{tabular}
\caption{Distance between best fit point and other dark energy models in $\gamma_{0}-\gamma_{a}$ and $r_{g0}-s_{g0}$ plane when $w_{0} =-0.95$ and $w_{a}=0$}. \label{dist1}
\end{center}
\end{table}
\noindent
It is clear that in $r_{g0}-s_{g0}$ phase plane the distances have always increased compared to $\gamma_{0}-\gamma_{a}$ or $w_{0}-w_{a}$ plane except for the GCG phantom model in the first case (where $w_{0}-w_{a}$ fixed). So we end up with the observation that in $r_{g0}-s_{g0}$ phase plane, the confidence contours are shrinked substantially whereas the distance between various dark energy models and the best fit point have also increased. This suggests that $r_{g0}-s_{g0}$ phase plane can have the potential to distinguish various dark energy models with future growth data.
\section{Conclusion}

\begin{table}[]
\begin{center}
\vspace{6pt}
\begin{tabular}{l|ccccc}
\hline\\
&  f(R) & DGP  & GCG-thawing  & GCG tracking  & GCG phantom  \\ \\
\hline 
\\
$w_{0}-w_{a}$ plane& 0.039 &0.151  &0.195 &0.181 &0.253  \\ \\
\hline \\
$r_{g0}-s_{g0}$ plane&0.163 &0.168  &0.410 &0.318 &0.905 \\
\\
\hline
\end{tabular}
\caption{Distance between best fit point and other dark energy models in $w_{0}-w_{a}$ and $r_{g0}-s_{g0}$ plane when $\gamma_{0} =0.545$ and $\gamma_{a}=0$}. \label{dist2}
\end{center}
\end{table}

To conclude, motivated by statefinder parameters $(r,s)$ for background expansion introduced by Sahni et al. \cite{Sahni03}, we introduce a similar pair of parameters $(r_{g}, s_{g})$ for the growth history of Universe. We call them {\it growth diagnostics}. Interestingly, both of these parameters take fixed value $(1,1)$ at all red shifts for $\Lambda$CDM model and for other dark energy behaviour they show very distinct trajectories in the $r_{g}-s_{g}$ phase plane. This property of the parameters is exploited to distinguish different dark energy models which produces nearly similar growth rate at present. This can be extremely useful to remove the degeneracies between different dark energy models as well as constraining them using future measurements like EUCLID. Using the EUCLID forecast data\cite{Amendola}, we show that the figure of merit for $r_{g0}-s_{g0}$ is almost thrice as compared to the figure of merit for $\gamma_{0}-\gamma_{a}$ or $w_{0}-w_{a}$. At the same time, individual models for late time acceleration of the Universe (invloving both dark energy as well as modified gravity models) are also more separated from the fiducal model in the $rg_{0}-sg_{0}$ plane, compared to the $\gamma_{0}-\gamma_{a}$ or $w_{0}-w_{a}$ plane. This shows that the $rg_{0}-sg_{0}$ plane can be more constraining in distinguishing various approaches to explain the late time acceleration of the Universe with future data from surveys like EUCLID.

\section{Acknowledgement} Authors would like to thank Cinzia di Porto for useful discussions. The author SA acknowledges SERC, Dept. of Science and Technology, Govt. of India for the financial support through the grant SR/S2/HEP-43/2009. The author AAS acknowledges the  partial support by the same grant.

\end{document}